\documentclass[aps,prd,floatfix,superscriptaddress,twocolumn]{revtex4}

\preprint{JLAB-THY-11-1447}

\usepackage{amsmath}
\usepackage{bm}
\usepackage{graphicx}

\begin{document}

\title{Constraints on the large-\textit{x} \textit{d/u} ratio
	from electron--nucleus scattering at \textit{x} $>$ 1}  

\author{O.~Hen}
\affiliation{Tel Aviv University, Tel Aviv 69978, Israel}

\author{A.~Accardi}
\affiliation{Hampton University, Hampton, Virginia 23668, USA}
\affiliation{Jefferson Lab, Newport News, Virginia 23606, USA}

\author{W.~Melnitchouk}
\affiliation{Jefferson Lab, Newport News, Virginia 23606, USA}

\author{E.~Piasetzky}
\affiliation{Tel Aviv University, Tel Aviv 69978, Israel}


\begin{abstract}
Recently the ratio of neutron to proton structure functions $F_2^n/F_2^p$
was extracted from a phenomenological correlation between the strength
of the nuclear EMC effect and inclusive electron--nucleus cross section
ratios at $x > 1$.  Within conventional models of nuclear smearing, this
``in-medium correction'' (IMC) extraction constrains the size of nuclear
effects in the deuteron structure functions, from which the neutron
structure function $F_2^n$ is usually extracted.  The IMC data determine
the resulting proton $d/u$ quark distribution ratio, extrapolated to
$x = 1$, to be $0.23 \pm 0.09$ with a 90\% confidence level. This is well 
below the SU(6) symmetry limit of 1/2 and significantly above the scalar 
diquark dominance limit of 0.
\end{abstract}

\maketitle

\section{Introduction}

Currently uncertainties in parton distribution functions (PDFs) at large
parton momentum fractions $x$ represent one of the main impediments to
the determination of the longitudinal structure of the nucleon in terms
of its fundamental constituents. The large-$x$ region provides a unique
opportunity for studying the flavor and spin dynamics of quarks in the
nucleon, with the $d/u$ quark distribution ratio in particular being
very sensitive to different mechanisms of spin-flavor symmetry breaking
\cite{MT96, Holt10}. Knowledge of PDFs at large $x$ is important also
for several other reasons, such as the reliable calculation of QCD
backgrounds in new physics searches at hadron colliders, especially
at large rapidities, as well as in neutrino oscillation experiments.

The systematics of uncertainties in parton distributions at large $x$
has been the focus of recent dedicated global QCD analyses by the
CTEQ-Jefferson Lab (CJ) Collaboration \cite{CTEQ6X, CJ11}, which
investigated the sensitivity of PDFs to different treatments of
nuclear corrections in deep-inelastic scattering (DIS) from deuterium.
While proton DIS data place strong constraints on the $u$-quark
distribution, neutron structure functions are needed in order to
determine also the $d$-quark PDF.  The absence of free neutron targets,
however, means that deuterium DIS data must be used to infer information
about the structure of the free neutron.

Uncertainties in the nuclear corrections in the deuteron, such as
those associated with nucleon off-shell effects and the large-momentum
components of the deuteron wave function, give rise to significant
uncertainties in the resulting $d/u$ ratio for $x \gtrsim 0.5$
\cite{CJ11}.  This prevents drawing any firm conclusions about the
$x \to 1$ behavior of $d/u$ predicted in various nonperturbative
and perturbative models, which range from 0 in models with scalar
diquark dominance \cite{Feynman72, Close73, Carlitz75} to $\approx 0.2$
in models with admixtures of axial-vector diquarks \cite{Cloet05} or
those based on helicity conservation \cite{FJ75}, and up to 0.5 in
models with SU(6) spin-flavor symmetry \cite{Close79}.

A recent analysis of the strength of the EMC effect in nuclei and
data on inclusive electron--nucleus scattering at $x > 1$ proposed
a phenomenological, theory-independent, determination of the neutron
to proton structure function ratio $F_2^n/F_2^p$, known as the
``in-medium correction'' (IMC) extraction \cite{IMC}.  The IMC
analysis is based on the observed correlation between the strength
of the nuclear EMC effect at intermediate $x$ ($x \approx 0.3-0.7$)
and the number of short range correlated nucleon--nucleon pairs in
light, medium, and heavy nuclei, which is then extrapolated to a
free nucleon.

In this report we combine the phenomenology of these two approaches,
and illustrate how the IMC extracted neutron structure function can in
principle limit the range of parameters describing nuclear corrections
in the deuteron, thereby significantly reducing the uncertainties in the
resulting $d/u$ ratio at large $x$.  As we shall see, the IMC analysis
favors values of $d/u$ at the upper end of the uncertainty band obtained
in the CJ global QCD fit \cite{CJ11}, indicating the presence of
significant nucleon off-shell corrections in the deuteron structure
function.

\section{Nuclear effects in the deuteron}

In the conventional description of DIS from the deuteron at $x \gg 0$,
the scattering is assumed to take place incoherently from individual
nucleons in the deuterium nucleus \cite{Jaffe85}.  In the weak binding
approximation (WBA), the deuteron structure function can be written as
a convolution of the bound nucleon structure functions $F_2^N$ and a
momentum distribution function of nucleons in the deuteron (also known
as the ``smearing function'') \cite{KPW94, KP06, Mel10}.

In Ref.~\cite{KP06}, Kulagin and Petti used a simple quark spectral
model to obtain a physically motivated parametrization of the nucleon
off-shell corrections (see also Refs.~\cite{KPW94, KMPW95, Liuti92,
MSTprd}).  The off-shell corrections were estimated by integrating
the quark--nucleon spectral function over quark virtualities up to
some high-momentum scale $\Lambda$ that depends on the nucleon
off-shell mass $p^2 = p^2_0 - \bm{p}^2 \not= M^2$, where
$p_0 = M_d - \sqrt{M^2 + \bm{p}^2}$ is the nucleon energy and $\bm{p}$
its momentum, with $M$ and $M_d$ the nucleon and deuteron masses.
Taking $\Lambda$ to be inversely proportional to the quark confinement
radius $R$ in the nucleon, its dependence on $p^2$ can be related
to the change in the size of the nucleon in the nuclear medium
(``nucleon swelling'').
The change of scale $\delta R$ and nucleon virtuality can be
conveniently parametrized in terms of a single parameter $\lambda$,
given by \cite{KP06}
\begin{eqnarray}
\lambda
&=& \left. { \partial\Lambda^2 \over \partial\log p^2 }
    \right|_{p^2=M^2}\
 =\ -2 {\delta R \over R} {\delta p^2 \over M^2}\, ,
\label{eq:lambda}
\end{eqnarray}
where $\delta p^2$ is the average nucleon virtuality ($p^2-M^2$)
in the deuteron.

The parameter $\lambda$ was chosen in Ref.~\cite{CJ11} to reproduce
the phenomenological values of the change of confinement radius from
the study of the nuclear EMC effect in the $Q^2$-rescaling model
\cite{CJRR}, $\delta R/R = 1.5\% - 1.8\%$.  This was somewhat
smaller than the nuclear-averaged value of $\delta R/R \approx 9\%$
obtained by fitting the off-shell correction to ratios of nuclear
structure functions for a range of nuclei \cite{KP06}.
While it is generally accepted that some off-shell corrections to the
convolution approximation are needed in order to describe nuclear
structure functions at large $x$ \cite{EMCrev}, their magnitude varies
considerably between different models \cite{KP06, Liuti92, MSTprd,
FS85, MSTplb, MSS97}, and on the definition of the smearing function.
(In fact, in some approaches such as the light-front \cite{FS81, Ma93,
ACHL09} explicit off-mass-shell corrections do not appear at all,
their effects instead being subsumed by higher Fock state components
or contact interactions.)

In the present analysis we treat $\lambda$ as a free parameter,
allowing it to be determined by the IMC extraction data for a given
virtuality $\delta p^2$.  The latter is computed from several modern
deuteron wave functions which give high-precision fits to
nucleon-nucleon scattering data, namely, the CD-Bonn \cite{CDBonn},
AV18 \cite{AV18}, and the relativistic WJC-1 and WJC-2 wave functions
\cite{WJC}, yielding values of the nucleon virtuality of
$\delta p^2/M^2 = -3.7\%, -4.5\%, -6.2\%$ and $-4.9\%$, respectively.
(The older Paris deuteron wave function \cite{Paris} gives a value
$\delta p^2/M^2 = -4.3\%$, similar to the AV18 model.)
This ``modified Kulagin-Petti'' (mKP) parametrization of the off-shell
corrections (\ref{eq:lambda}) allows a wide range of models to be
assessed in terms of a single parameter, the nucleon swelling
$\delta R/R$, for a given deuteron wave function.

\begin{figure}[h]
\includegraphics[width=9cm]{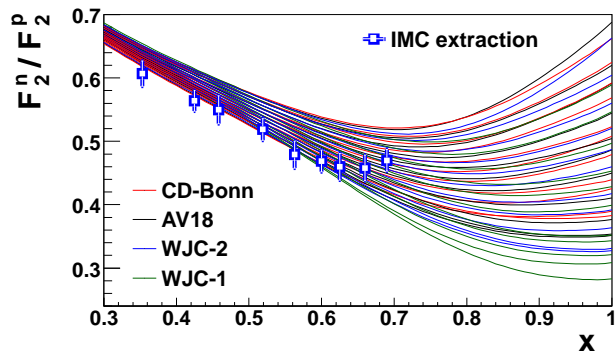}
\caption{Neutron to proton structure function ratio $F_2^n/F_2^p$ from
	the CJ global QCD fit \cite{CJ11} assuming different deuteron
	wave functions:
	CD-Bonn (red), AV18 (black), WJC-1 (green) and WJC-2 (blue).
	The curves correspond to different nucleon swelling levels,
	$\delta R/R$, ranging from 0\% (lowest curves) to 3\%
	(highest curves) in steps of 0.3\%.  The IMC data (squares)
	\cite{IMC} and the fits are at a fixed $Q^2 = 12$~GeV$^2$.}
\label{fig:npratio}
\end{figure}

\section{IMC constraints on the $\bm{d/u}$ ratio}

In Fig.~\ref{fig:npratio} the ratio of neutron to proton structure
functions $F_2^n/F_2^p$ at $Q^2 = 12$~GeV$^2$ is shown for various
deuteron wave functions and swelling levels $\delta R/R$, ranging
from 0\% to 3\%, in increments of 0.3\%, using the WBA smearing
function and the mKP off-shell model.  For each combination of wave
function and swelling parameters, the structure functions are computed
from the CJ global next-to-leading order QCD fit of PDFs as described
in Ref.~\cite{CJ11}, using a flexible parametrization for the $d$-quark
PDF which allows finite $d/u$ values as $x \to 1$.
Each of the fitted PDF sets represented by the curves in
Fig.~\ref{fig:npratio} gives a similar quality fit to the global data
base used in Ref.~\cite{CJ11}, by allowing the changes in the nuclear 
corrections to the deuteron $F_2^d$ structure function to be compensated
by corresponding changes in the $d$-quark PDF (inducing a similar change
in the calculated neutron $F_2^n$).  The curves are compared with the
$F_2^n/F_2^p$ ratios obtained from the IMC extraction over the range
$0.35 \lesssim x \lesssim 0.7$.

\begin{figure*}[ht]
\includegraphics[width=8cm]{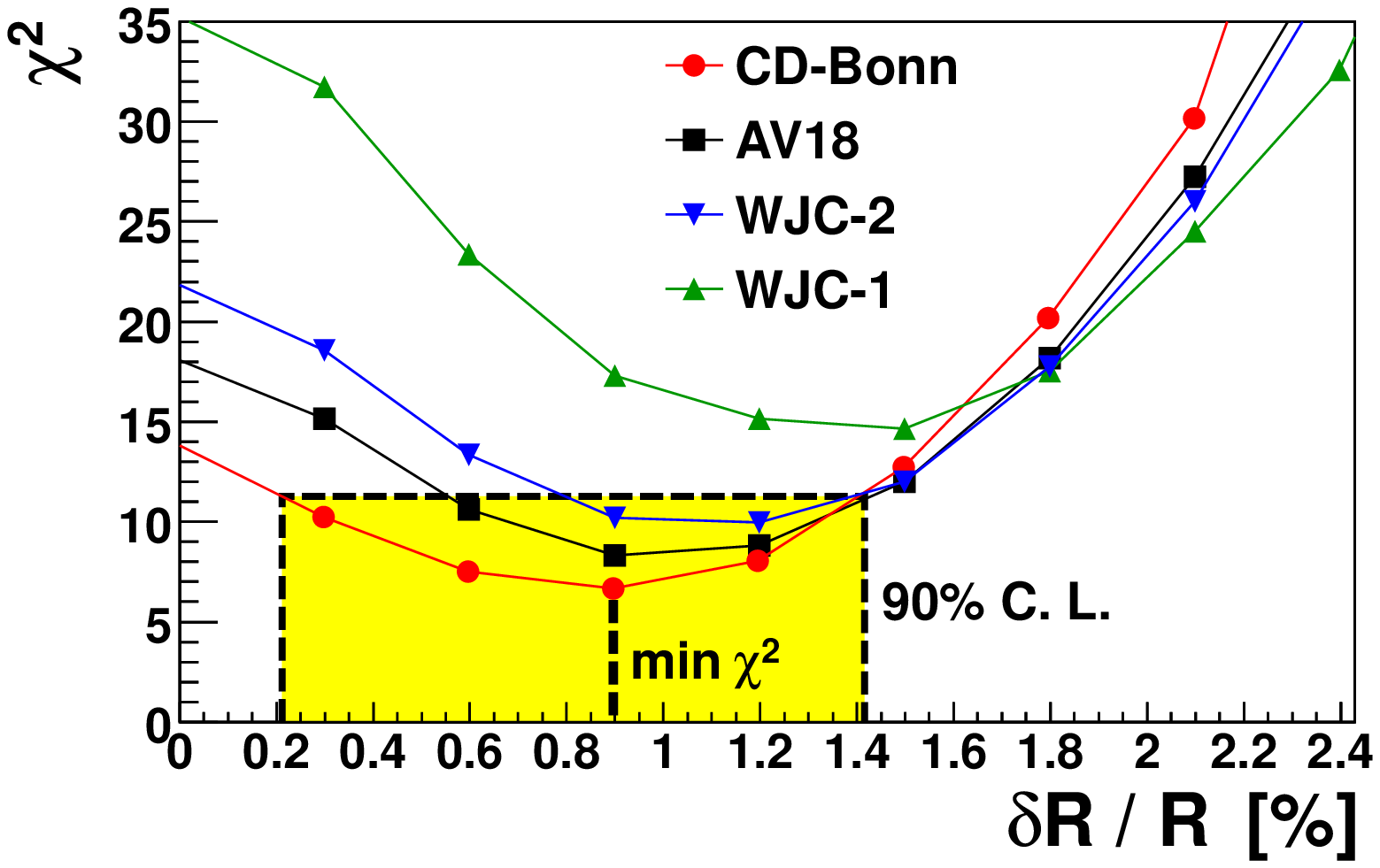}
\includegraphics[width=8cm]{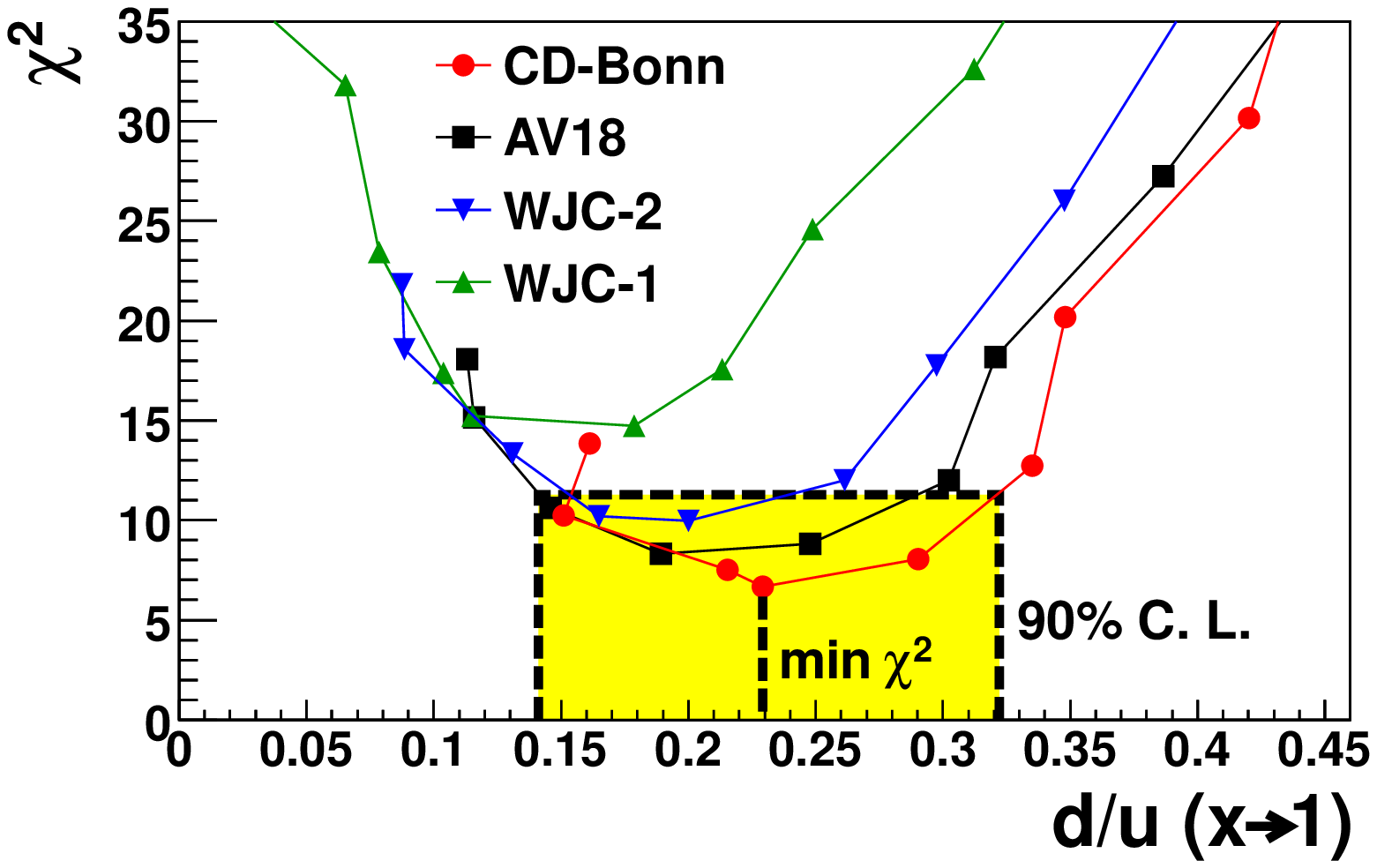}
\caption{Total $\chi^2$ for fits of the calculated $F_2^n/F_2^p$
	ratios in Fig.~\ref{fig:npratio} to the IMC extraction
	data \cite{IMC} for various deuteron wave functions
	(CD-Bonn -- circles, AV18 -- squares, WJC-2 -- inverted
	triangles, WJC-1 -- triangles), as a function of the swelling
	level $\delta R/R$ {\it (left)}, and the $d/u$ ratio in	the
	$x \to 1$ limit {\it (right)}.  The 90\% confidence levels
	are indicated by the shaded (yellow) box, and the minimum
	$\chi^2$ values by the vertical dashed line.}
\label{fig:chi2}
\end{figure*}

To constrain the nuclear correction uncertainty in $F_2^n/F_2^p$, we    
calculate the $\chi^2$ of the IMC data for each deuteron wave function
and swelling combination.  This is shown in Fig.~\ref{fig:chi2}~(left)
as a function of the nucleon swelling $\delta R/R$ for the different
deuteron wave functions.  Note that the wave function determines not   
only the average nucleon virtuality $\delta p^2$ in the deuteron, but 
also the amount of binding and Fermi motion in the smearing function   
\cite{CJ11, Mel10}.
For the choice of confidence level (C.~L.) we treat the deuteron wave
function as a (discrete) parameter, and consider a 90\% C.~L. for two
free parameters, corresponding to an increase in $\chi^2$ of 4.61 above
the minimum.
With this C.~L. the IMC extraction constrains the swelling levels 
to the range $\delta R/R = 0.2\% - 1.4\%$, with a preference for the
CD-Bonn, AV18 and WJC-2 wave functions.  The minimum $\chi^2$ occurs  
for the CD-Bonn model at $\delta R/R = 0.9\%$.  The minimum $\chi^2$  
for the WJC-1 wave function at $\delta R/R \approx 1.5\%$ lies outside
of the 90\% C.~L. and is disfavored by the IMC data.

The implications of these constraints for the $d/u$ ratio in the
limit $x \to 1$ are illustrated in Fig.~\ref{fig:chi2}~(right), 
where the $\chi^2$ is shown as a function of the limiting $d/u$ value
of each PDF fit.  The IMC extraction yields a $d/u$ limiting value of
$0.23 \pm 0.09$ at the 90\% C.~L. (at the 99\% C.~L., using a $\chi^2$
increase of 9.21, the uncertainty would increase to $\pm 0.13$).
These results strongly disfavor the SU(6) value $d/u = 1/2$, as well as
the $d/u \to 0$ limit predicted in models with scalar diquark dominance.
Furthermore, global PDF analyses often assume the same functional $x$
dependence for both $u$ and $d$ quark distributions, forcing the $d/u$
ratio to approach either zero or infinity in the $x \to 1$ limit.
The results shown in Fig.~\ref{fig:chi2}, however, suggest that a
more flexible parametrization for the $d/u$ ratio, which allows
finite $x \to 1$ limits, may be more realistic \cite{CJ11}.

\begin{figure}[b]
\includegraphics[width=9cm]{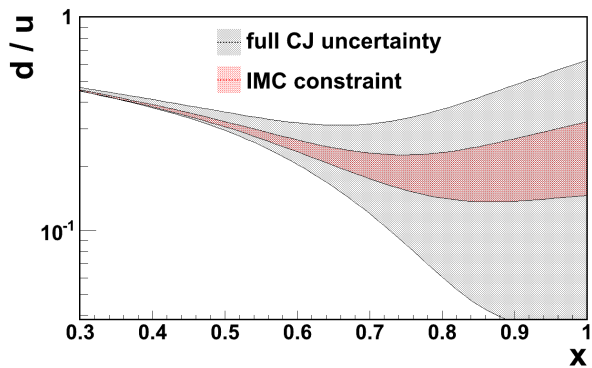}
\caption{$d/u$ ratio at $Q^2 = 12$~GeV$^2$ with the full theoretical
	uncertainty from Ref.~\cite{CJ11} (black) and with the IMC
	constraint at the 90\% C.~L. (red).}
\label{fig:duratio}
\end{figure}

The resulting uncertainty bands on the $d/u$ ratio are shown in
Fig.~\ref{fig:duratio}, including the full theoretical uncertainty
from the CJ global fit \cite{CJ11}, and the 90\% C.~L. extracted
from the IMC constraints.  Even though the IMC extraction only
covers an $x$ range of $\approx 0.35-0.7$, it nevertheless imposes a
tight constraint on the $d/u$ parton distributions ratio for $x \to 1$.

\section{Summary}

Within a global PDF analysis we have studied the constraints imposed by
the theory-independent IMC-extracted $F_2^n$ structure function data on
nuclear corrections in deuterium.  These phenomenologically extracted
data strongly support the presence of off-shell modifications of nucleons
in the deuteron, and constrain their magnitude to a more limited range
than in the recent CJ global QCD analysis without the IMC data
\cite{CJ11}.  The IMC data also disfavor deuterium wave functions with
very ``hard'' momentum distributions, such as for the \mbox{WJC-1}
nucleon-nucleon potential \cite{WJC}, which produce a shallow EMC ratio
$F_2^d/F_2^N$ at intermediate and large $x$ \cite{CJ11, Mel10}.

While the $u$-quark PDF is well constrained by the proton DIS data,
the lack of a free neutron target makes the $d$-quark distribution very
sensitive to the assumptions used to calculate the nuclear correction
in the deuteron.  The use of the IMC-extracted neutron structure
function directly constrains the $d$-quark PDF for $x \lesssim 0.7$,
and indirectly for $x \to 1$.  We find the $d/u$ ratio in the limit
$x \to 1$ to be $0.23 \pm 0.09$ at the 90\% confidence level, in
agreement with models predicting intermediate values of $d/u$ between
the SU(6) symmetry and scalar diquark dominance limits
\cite{FJ75, Cloet05}.

Of course, these conclusions strongly depend on the assumptions
underlying the IMC extraction of $F_2^n$ \cite{IMC}.  Some of these are
being tested through the study of DIS events with a tagged high-momentum
proton recoil at Jefferson Lab \cite{DEEPS}, and will be the subject of
a similar experiment at the future 12~GeV upgraded facility
\cite{E12-11-107}.  The ultimate arbiter, however, will be data on
free or nearly free neutron targets, such as from the BoNuS experiment
\cite{BONUS} at Jefferson Lab that collected DIS data up to
$x \approx 0.6$, or its future 12~GeV extension \cite{BONUS12} that
will reach $x \approx 0.8$.  Further avenues to direct experimental
constraints on $d/u$ at large $x$ include the 12~GeV MARATHON experiment
\cite{MARATHON} at Jefferson Lab on DIS from the $^3$He--$^3$He mirror
nuclei and the parity-violating DIS program on a hydrogen target
\cite{SOLID}, as well as the measurement of $W$ boson asymmetries
at large rapidities in $p\bar p$ collisions at the Tevatron or in
$pp$ scattering at RHIC and the LHC \cite{CJ11, BAMO}.

\section*{Acknowledgements}

We are grateful for collaboration and many fruitful discussions with
D.~W.~Higinbotham, J.~F.~Owens and L.~B.~Weinstein.
This work was supported by the Israel Science Foundation, the US-Israeli
Bi-National Science Foundation, the US DOE contract No.~DE-AC05-06OR23177,
under which Jefferson Science Associates, LLC operates Jefferson Lab,
and the US National Science Foundation under NSF Awards No.~1002644.


\end{document}